\documentclass[aps,prl,superscriptaddress,twocolumn,floatfix,showpacs]{revtex4-1}
\usepackage{graphicx}  % needed for figures
\usepackage{dcolumn}   % needed for some tables
\usepackage{bm}        % for math
\usepackage{amssymb}   % for math
\usepackage{color}
\usepackage{pdfpages}
\usepackage{amsmath}   % for nonumber in equation
\usepackage{comment}
\begin{document}

\title{Formation of Non-reciprocal Bands in Magnetized Diatomic Plasmonic Chains}

\author{C. W. Ling}\affiliation{Department of Applied Physics, The Hong Kong Polytechnic University, Hong Kong}
\author{Jin Wang}\affiliation{Department of Physics, Southeast University, Nanjing, China}
\author{Kin Hung Fung}\email{khfung@polyu.edu.hk}\affiliation{Department of Applied Physics, The Hong Kong Polytechnic University, Hong Kong}

\date{\today}

%old abstract
%\begin{abstract}It has been shown that plasmonic chains, an array of spatial periodic metal nanoparticles, support energy transmission via coupled surface plasmon modes. The coupled modes are characterized by dispersion relation $\omega(k)$, and are normally symmetric $\omega(-k)=\omega(k)$ (reciprocal bands). Spectral reciprocity is known to be protected by either time reversal ($\mathcal{T}$) or inversion ($\mathcal{P}$) symmetry. However, breaking of $\mathcal{P}$ and $\mathcal{T}$ symmetries are not sufficient to achieve $\omega(-k)\neq\omega(k)$. We use a magnetized plasmonic diatomic chain system to show that breaking pi-rotation time-reversal ($\mathcal{RT}$) symmetry is also a necessary condition. By applying point dipole approximation and eigendecomposition method, we show that the dispersion relation of the chain enables one-way propagation with applications like optical isolators and switches. Optical transmission and hybridization of bands are also studied and demonstrated. Explicit form of symmetry operators are also deduced and given in the paper.\end{abstract}

\begin{abstract}
We show that non-reciprocal bands can be formed in a magnetized periodic chain of spherical plasmonic particles with two particles per unit cell. Simplified form of symmetry operators in dipole approximations are used to demonstrate explicitly the relation between spectral non-reciprocity and broken spatial-temporal symmetries. Due to hybridization among plasmon modes and free photon modes, strong spectral non-reciprocity appears in region slightly below the lightline, where highly directed guiding of energy can be supported. The results may provide a clear guidance on the design of one-way waveguides.
\end{abstract}

%73.20.Mf Collective excitations
%78.67.Pt Multilayers; superlattices; photonic structures; metamaterials
%11.30.Qc Symmetry breaking

\maketitle
%%%%%%introduction%%%%%%
\section{I. Introduction}

Breaking Lorentz reciprocity~\cite{jakong1970,caltman1982} in optics has been of great interest to physicists for many decades. In recent years, the asymmetry in dispersion relation $\omega(-\bf{k})\neq\omega(\bf{k})$ (i.e, spectral non-reciprocity)~\cite{recamley1987, afigotin2001} has drawn a lot of interest because of its possible topological nature~\cite{tochiai2015,tochiai2015} and potential applications such as on-chip optical isolators, unidirectional waveguides, and circulators~\cite{fdmhaldane2008,kfang2011,zwang2008,hiroyuki2008}. Non-reciprocal bands predicted by topological band theory usually appears as surface modes attached to two-dimensional (2D) or three-dimensional (3D) bulk photonic systems. To make the device more compact, approaches based on symmetry breaking in waveguide structures usually suggest complex geometries such as helical structures~\cite{yhadad2010,achristofi2013,tochiai2015,pjcheng2015}.

Spectral reciprocity, $\omega(-\bf{k})=\omega(\bf{k})$, can be protected by time reversal symmetry ($\mathcal{T}$) and spatial symmetries such as inversion ($\mathcal{P}$) \cite{recamley1987,afigotin2001}, in addition to the local symmetries in permitivity or permeability tensors (${\bm{\epsilon}}^{\rm{T}}={\bm{\epsilon}}$ or ${\bm{\mu}}^{\rm{T}}={\bm{\mu}}$). It is easy to understand that $\mathcal{T}$ symmetry can be broken by external static magnetic field \cite{tochiai2015}, while $\mathcal{P}$ symmetry can be broken by using asymmetric structures such as chiral structures \cite{yhadad2010,achristofi2013,tochiai2015,pjcheng2015} or a symmetrical structure under external magnetic field of specific orientation \cite{xlin2013}. However, spectral reciprocity can also be protected by a combination of symmetries such as spatial-temporal symmetries, which add more complexities in the design of non-reciprocal waveguides.

In this paper, we use compact non-chiral magnetized plasmonic waveguides consisting of only spherical particles to demonstrate how spectral reciprocity can be protected by a rotation-time-reversal ($\mathcal{RT}$) symmetry (i.e., time reverse followed by rotation of $180^\circ$ about propagation $x$-axis). In the $\mathcal{RT}$ symmetry broken case, we show that asymmetric dispersion relation can be supported. By coupling the hybridized bands with light lines, this simple system further supports one-way wave propagation and energy transmission within a finite range of frequencies.

\begin{figure}[htbp]
\centering
\includegraphics[width=2.5in]{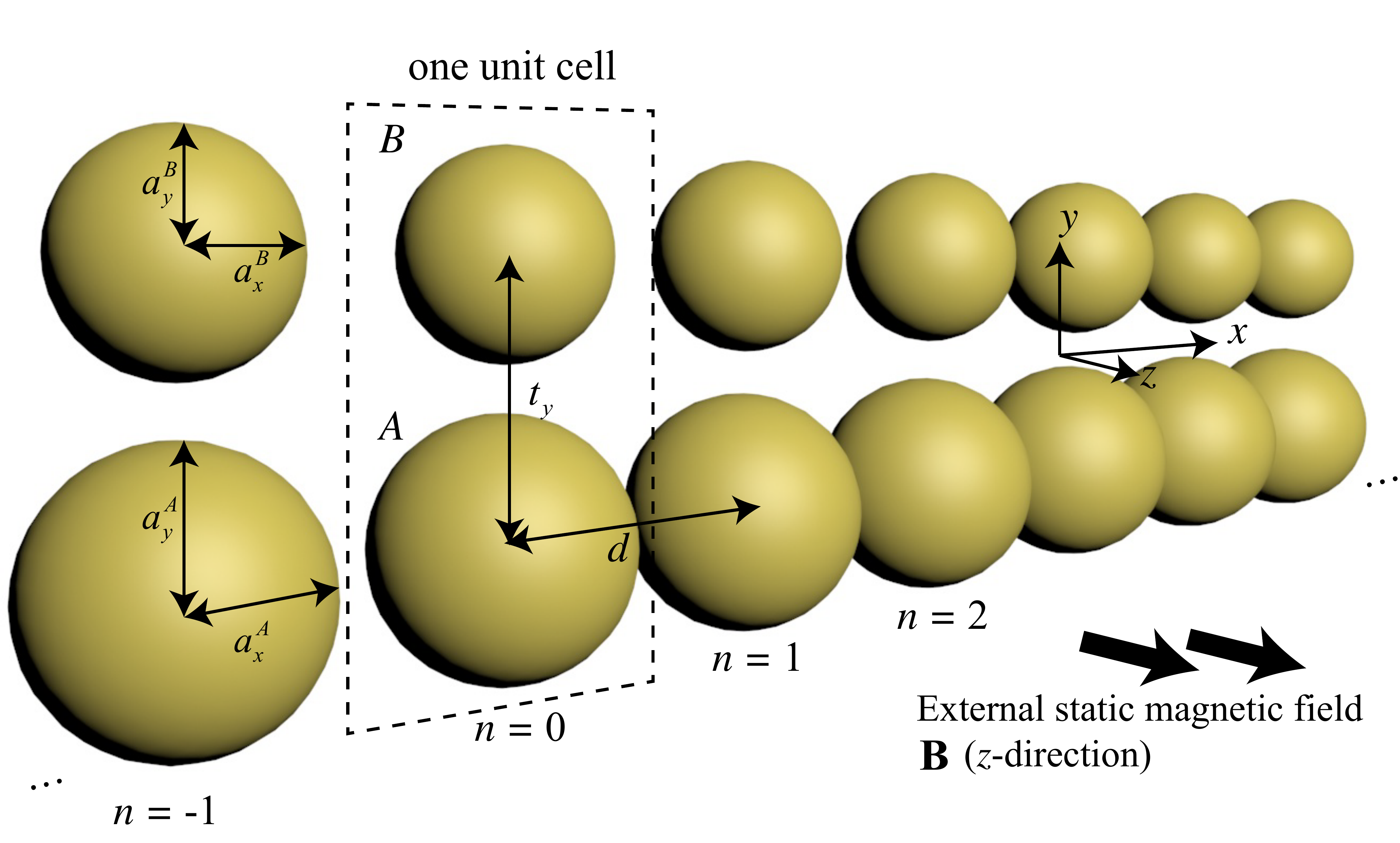}
\caption{\small(Color online) Geometry of a diatomic chain system. The chain contains two metallic nanoparticles in a unit cell, denoted by $A$ and $B$. They can be elliptical or spherical, dependent on the values of semi-major axis $a^{\sigma}_x$ and semi-minor axis $a^{\sigma}_y$, where $\sigma=A$ or $B$. In the case of spherical, $a^{\sigma}_x=a^{\sigma}_y$. $A$ and $B$ are separated with distance $t_y$. External static magnetic field $\bf{B}$ is applied in $z$ direction. Length of a unit cell is $d$. The chain breaks $\mathcal{T}$, $\mathcal{P}$, and $\mathcal{RT}$ symmetry.}
\label{fig:01}
\end{figure}

\begin{figure*}[htbp]
\centering
\includegraphics[width=6.4in]{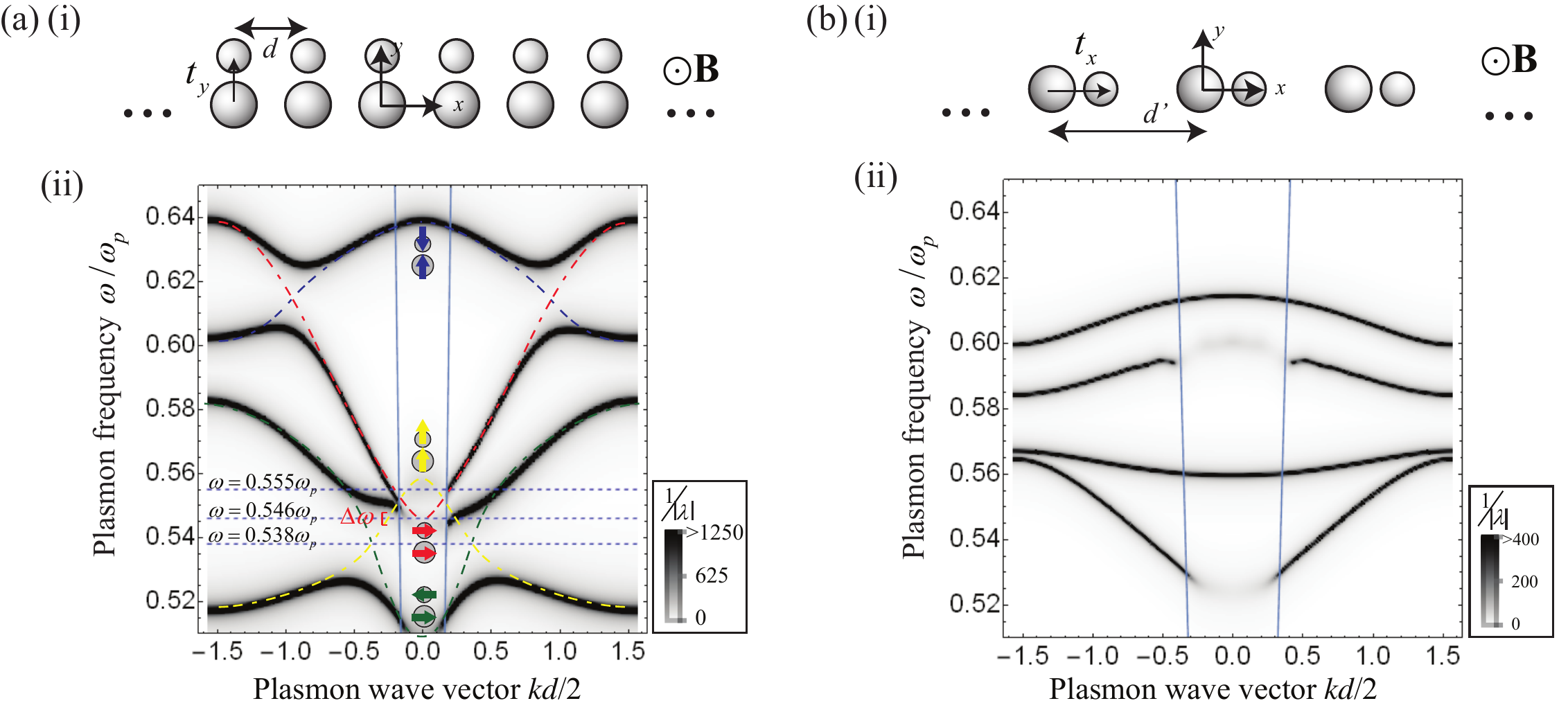}
\caption {\small(Color online) Dispersion relations for an infinitely long and magnetized diatomic chain system without $\mathcal{RT}$ symmetry (a) and with $\mathcal{RT}$ symmetry (b). Sub-figure (i) is the chain geometry, while (ii) is corresponding dispersion relation obtained by density plot of $1/\left|\lambda\right|$. (a)(ii) shows non-reciprocal bands ($\omega(k)\neq \omega(-k)$), while only reciprocal bands ($\omega(k)=\omega(-k)$) could be seen in (b)(ii). Bands related to $z$ component are separated and not shown here. The blue solid lines are dispersions of free photon modes. Inner particle separation in (a) is $t_y=0.75d$, and in (b) is $t_x=0.425d'$, where $d'=2d$. Furthermore, (a)(ii) can be understood as hybridization of 4 bands, which are schematically drawn in dashed for guidance. Each dashed line represents different oscillation modes at $k = 0$, which is labeled by arrows at the middle. $\Delta \omega$ is the range at which only one-way propagation modes are allowed.}
\label{fig:02dispersion}
\end{figure*}

\section{II. Model and methods}
%coupled dipole equation
We start by considering a magnetized diatomic system as shown in Fig.~\ref{fig:01}. The chain contains two types of metallic nanoparticles with different sizes, namely particle $A$ and $B$, with the same dielectric constant ${\bm{\epsilon}}(\omega)/\epsilon_0$. The two nanoparticles formed ``atoms" of a unit cell, and hence it is regarded as a dimer chain. As long as the nanoparticles are not too close together, the electromagnetic responses of the nanoparticles can be modeled by electric dipoles~\cite{whweber2004}. We denote the dipole moment of nanoparticle in the $n$th unit cell as ${\bf{p}}_{n;\sigma}$ where $\sigma=A$ or $B$ for type $A$ and $B$ particles, respectively. These dipole moments satisfy a set of self-consistent equations, known as the coupled dipole equations \cite{whweber2004,khfung2007,cwling2010}:
\begin{equation} \label{eq:01coupled}
\sum\limits_{m,\sigma'}{
    \left({\bm{\alpha}}^{-1}_{\sigma}\delta_{nm}\delta_{\sigma\sigma'}
    -{\bf{G}}_{nm\sigma\sigma'}\right)
    }
{\bf{p}}_{m;\sigma'}={\bf{E}}^{\rm{ext}}_{n;\sigma},
\end{equation}
in which $m$ runs from $-N$ to $N$, $\sigma'=A$ or $B$, $\delta_{nm}$ is Kronecker delta function, and ${\bf{E}}^{\rm{ext}}_{n;\sigma}$ is the external driving field. We note that ${\bm{\alpha}}_{\sigma}$ is the quasi-static polarizability with radiation correction of nanoparticle $\sigma$ and ${\bf{G}}_{nm\sigma\sigma'}$ is interaction between dipoles ${\bf{p}}_{m;\sigma'}$ and ${\bf{p}}_{n;\sigma}$. Expressions are given in Appendix A [see Eqs.~(\ref{eq:a1polarizability}) and (\ref{eq:a4green})].

%%in pk form for dispersion
%\subsection{Dispersion relation}
We first consider the case without external driving field (${\bf{E}}^{\rm{ext}}_{n;\sigma}=0$) and $N\rightarrow\infty$. Since the system is spatially periodic, by Bloch's theorem we can write ${\bf{p}}_{m;\sigma}={\bf{p}}_{k;\sigma}e^{ikmd}$, where $k$ is wave vector. This simplifies Eq.~(\ref{eq:01coupled}) into a 6 by 6 matrix form (see Eq.~(\ref{eq:a3coupled})):
\begin{equation}\label{eq:02matrixcouple}
\begin{array}{l}
    {\bf{M}}_k(\omega)
    \left[ {\begin{array}{*{20}{c}}
        {{{\bf{p}}_{k;A}}}  \\
        {{{\bf{p}}_{k;B}}}  \\
    \end{array}} \right]=0, \\
 \end{array}
\end{equation}
where
\begin{equation}\label{eq:03matrixoperator}
\begin{array}{l}
    {\bf{M}}_k(\omega)=
        \left[ {\begin{array}{*{20}{c}}
        {{\bm{\alpha}}_A^{ - 1}} & 0  \\
        0 & {{\bm{\alpha}}_B^{ - 1}}  \\
        \end{array}} \right]
        -
        \left[ {\begin{array}{*{20}{c}}
        {{{\bf{G}}_{kAA}}} & {{{\bf{G}}_{kAB}}}  \\
        {{{\bf{G}}_{kBA}}} & {{{\bf{G}}_{kBB}}}  \\
        \end{array}} \right].
 \end{array}
\end{equation}
In the above, ${\bf{G}}_{k\sigma\sigma'}=\sum\nolimits_{m}{\bf{G}}_{0m\sigma\sigma'}e^{ikmd}$, is the interaction between nanoparticles $\sigma$ and $\sigma'$ in $k$ space. Instead of solving $\det{\bf{M}}_k(\omega)=0$ to obtain the dispersion relation \cite{hfzhang2012,whweber2004,yhadad2010}, we may apply the eigen-response theory to evaluate the dispersion relation \cite{khfung2007,jwdong2013}. Similar to the eigen-response theory, we plot $1/\left|\lambda\right|$ as a function of $k$ and $\omega$, where $\left|\lambda \right|=\left|\lambda(k,\omega)\right|$ is the smallest absolute value of the eigenvalue of the matrix ${\bf{M}}_k(\omega)$. This quantity gives huge value when there is resonance, and is plotted in Fig.~\ref{fig:02dispersion}(a). Note that dynamic dipolar Green's function \cite{yhadad2010} is used, and the infinite series in the interaction up to $|m|=120$.

\section{III. Formation of nonreciprocal bands}
In Fig.~\ref{fig:02dispersion}(a), we showed the geometry and corresponding dispersion relation in (i) and (ii). This case considers nanoparticles are spherical with radius $a_x^A=0.35d$ and $a_x^B=\sqrt[3]{{0.5}}a_x^A$, inner particle separation $t_y=0.75d$, and plasma wavelength $\lambda_d\equiv c/(2\pi\omega_p)=10d$, where $\omega_p$ and $c$ are plasma frequency and light speed in vacuum. Cyclotron frequency $\omega_c=q|{\bf{B}}|/m=0.005\omega_p$, in which $\bf{B}$, $m$, and $q$ are external static magnetic field, electron mass, and electron charge. As $\omega_c\propto|{\bf{B}}|$, it is treated as a variable to indicate the magnitude of ${\bf{B}}$. Furthermore, for simplicity, simple lossless Drude model is used \footnote{Where plasma collision frequency $\gamma=0$.}, and bands related to $p_{k;\sigma}^z$, the $z$ component of ${\bf{p}}_{k;\sigma}$, are not shown. It is because ${\bf{B}}$ is in $z$ direction, so $p_{k;\sigma}^x$ are coupled with $p_{k;\sigma}^y$ but not $p_{k;\sigma}^z$, hence the bands can be separated. Note that the two solid blue lines are light lines, the dispersion of free photon modes. Region within light lines is light cone, modes in light cone are radiative and therefore not sustainable \cite{yhadad2013,yhadad2010,smwang2008}.

Figure \ref{fig:02dispersion}(a) shows a case with non-reciprocal (asymmetric) bands. For comparison, using the same formalism, a case with reciprocal (symmetric) bands is shown in Fig.~\ref{fig:02dispersion}(b). This is a case where nanoparticle $B$ and $A$ are on the same axis, and horizontal inner separation is $t_x=0.425d'$, where $d'$ is length of unit cell in (b). We set $d'=2d$, twice than that in (a), therefore the light cone in (b) is bigger than (a). Both (a) and (b) share the same nanoparticles $A$ and $B$, cyclotron frequency $\omega_c$, and plasma wavelength $\lambda_d$ ($\lambda_p=5d'$ in this case).

The non-reciprocal bands in Fig.~\ref{fig:02dispersion}(a)(ii) predicts only guided modes with positive group velocities are allowed within the range $\Delta \omega$. This results one-way propagation behavior, which can be utilized as an isolator. Note the operation frequency $\Delta \omega$ is relatively broad, for example, about 50 times wider than the structure suggested in Ref. \cite{yhadad2010}. Range of operation frequency is about $1\times10^{-4}\omega_p$ in Ref. \cite{yhadad2010}, while we have about $5\times10^{-3} \omega_p$. The isolator also has a lower requirement on the external magnetic field, where cyclotron frequency $\omega_c\sim0.005\omega_p$, is about 10 times smaller than that used in Refs. \cite{yhadad2010,achristofi2013}, and about 100 times smaller than that in Ref. \cite{zongfuyu2013}.

We notice that the non-reciprocal bands in Fig.~\ref{fig:02dispersion}(a) are obtained by means of the simultaneous violation of certain symmetries, $\mathcal{P}$, $\mathcal{T}$, and $\mathcal{RT}$ symmetries, whereas the only breaking of $\mathcal{P}$ and $\mathcal{T}$ still makes the system reciprocal shown in Fig.~\ref{fig:02dispersion}(b). We will discuss the relation between reciprocity and related symmetries below.

\subsection{A. Reciprocity protected by $\mathcal{P}$ and $\mathcal{T}$ symmetry}
%protected by P
Given ${\bf{p}}_{n;\sigma}(t)={\bf{p}}_{k;\sigma} e^{iknd-i\omega t}$ is a solution of a system with frequency $\omega$ and wave vector $k$. Spatially inverted state will be ${\mathcal{P}}\left[{\bf{p}}_{n;\sigma}(t)\right]={\mathcal{P}}\left({\bf{p}}_{k;\sigma}\right) e^{-iknd-i\omega t}$. $\mathcal{P}$ turns $(x,y,z)$ into $(-x,-y,-z)$, flips the direction of vector quantities, but does not modify $\omega$. If the system has $\mathcal{P}$ symmetry, the inverted state will also be the solution of the system. The factor $e^{-iknd-i\omega t}$ of the new solution means that it is a solution with frequency $\omega$ but wave vector $-k$. This tells us that we are always able to find a solution with frequency $\omega$ but wave vector $-k$ if the system has $\mathcal{P}$ symmetry, and thus the dispersion must be symmetric ($\omega(k)=\omega(-k)$).

\begin{figure}[htbp]
\centering
\includegraphics[width=3.3in]{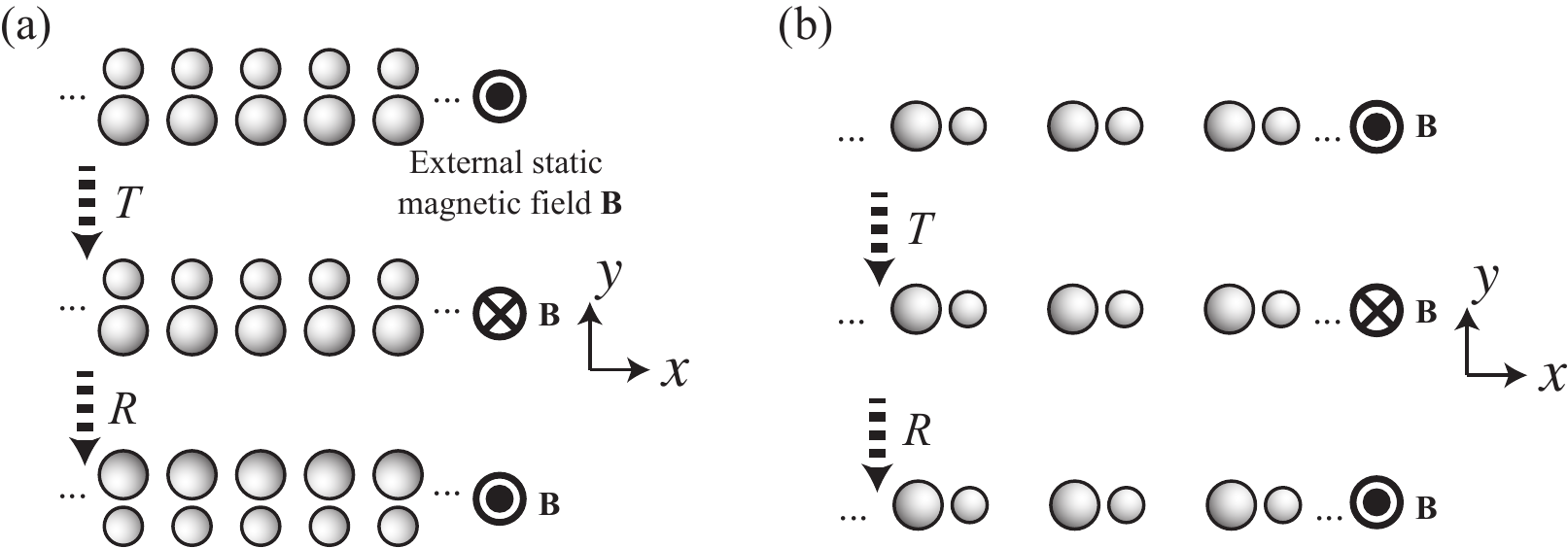}
\caption{\small $\mathcal{RT}$ acts on the diatomic chain systems shown in Fig.~\ref{fig:02dispersion}. $\mathcal{T}$ is time reversal operation, which flips the direction of $\bf{B}$ when acting on the system. This can be understood as the motion of electrons are reversed \cite{JDJ1999}. $\mathcal{R}$ is pi-rotation about $x$-axis. It exchanges the positions of $A$ and $B$ and flips the direction of $\bf{B}$ in case (a), only flips $\bf{B}$ in case (b). $\mathcal{RT}$ is the operation $\mathcal{T}$ followed by $\mathcal{R}$. For (a), as the final system is not identical to the initial one, the chain does not have $\mathcal{RT}$ symmetry. For (b), the final system is identical to the initial one, the chain has $\mathcal{RT}$ symmetry.}
\label{fig:03RT}
\end{figure}

%protected by T
Time reversed state is obtained by taking complex conjugate of the frequency component (see Ref.~\cite{ck2011}, also appendix E),
\begin{align}\label{eq:05Toperator}
{\mathcal{T}}\left[{\bf{p}}_{n;\sigma}(t)\right]
&= \left({\bf{p}}_{k;\sigma} e^{iknd}\right)^*e^{-i\omega t} \nonumber\\
&= {\bf{p}}_{k;\sigma}^* e^{-iknd-i\omega t}.
\end{align}
This will be another solution if the system has $\mathcal{T}$ symmetry. Similarly, it is a solution with frequency $\omega$ but wave vector $-k$, as it has the factor $e^{-iknd-i\omega t}$. Using arguments as those in $\mathcal{P}$ symmetry, we know the bands are symmetric as long as there is $\mathcal{T}$ symmetry. We see that, from Fig.~\ref{fig:03RT}, direction of external $\bf{B}$ will be flipped if $\mathcal{T}$ is operated on the system. Thus, the presence of external ${\bf{B}}$ breaks $\mathcal{T}$ symmetry, as the transformed one is not identical to the initial one.

Note that lossy system will also break $\mathcal{T}$ symmetry. We assumed the material is lossless, and the effect of radiation loss is compensated if we count the contributions from all nanoparticles in the infinite chain system \cite{Andrea2006}. So the $\mathcal{T}$ symmetry will be broken by external $\bf{B}$ only.

\subsection{B. Reciprocity protected by $\mathcal{RT}$ symmetry}
%protected by RT
$\mathcal{R}$ is pi-rotation operator, which rotates the system with $180^\circ$ about $x$-axis. In the case of diatomic chain system shown in Fig.~\ref{fig:01}, the direction of external ${\bf{B}}$ is flipped, and positions of $A$ and $B$ are exchanged, see Fig.~\ref{fig:03RT}. $\mathcal{RT}$ is the operation $\mathcal{T}$ followed by $\mathcal{R}$.
When $\mathcal{RT}$ acts on the solution, we have
\begin{align}\label{eq:07RT}
{\mathcal{RT}}\left[{\bf{p}}_{n;\sigma}(t)\right]
&= {\mathcal{R}}\left({\bf{p}}_{k;\sigma}^* e^{-iknd-i\omega t}\right) \nonumber\\
&= {\mathcal{R}}\left({\bf{p}}_{k;\sigma}^*\right) e^{-iknd-i\omega t},
\end{align}
here we used the property that $\mathcal{R}$ is not related to position $x$ and time $\mathcal{T}$. Again, if the system has $\mathcal{RT}$ symmetry, Eq.~(\ref{eq:07RT}) will be a solution with frequency $\omega$ but wave vector $-k$. This shows that the bands are symmetric about $k$.

The matrix representation of $\mathcal{R}$ depends on system geometry. For the chain system shown in Fig.~\ref{fig:02dispersion}(a), $\mathcal{R}$ not only exchanges the positions of $A$ and $B$, but also rotates the vector quantities when acting on the states. Denoting the $3\times3$ rotation matrix which rotates a vector about $x$-axis with $180^\circ$ by ${\bf{R}}={\rm{diag}}(1,-1,-1)$, then the $\mathcal{RT}$ transformed state is:
\begin{subequations}
\begin{equation}\label{eq:07RTfig1}
{\mathcal{RT}}\left[{{\bf{p}}_{n;\sigma }}(t)\right] = \left[ {\begin{array}{*{20}{c}}
   0 & {\bf{R}}  \\
   {\bf{R}} & 0  \\
\end{array}} \right]\left[ {\begin{array}{*{20}{c}}
   {{\bf{p}}_{k;A}^*}  \\
   {{\bf{p}}_{k;B}^*}  \\
\end{array}} \right]{e^{ - iknd - i\omega t}}.
\end{equation}

For the chain system shown in Fig.~\ref{fig:02dispersion}(b), $\mathcal{R}$ does not exchange positions of $A$ and $B$, so the $\mathcal{RT}$ transformed state is
\begin{equation}\label{eq:07RTfig4}
{\mathcal{RT}}\left[ {{{\bf{p}}_{n;\sigma }}(t)} \right] = \left[ {\begin{array}{*{20}{c}}
   {\bf{R}} & 0  \\
   0 & {\bf{R}}  \\
\end{array}} \right]\left[ {\begin{array}{*{20}{c}}
   {{\bf{p}}_{k;A}^*}  \\
   {{\bf{p}}_{k;B}^*}  \\
\end{array}} \right]{e^{ - iknd - i\omega t}}.
\end{equation}
\end{subequations}
%%examples on RT plasmonic chains

For the system of Fig.~\ref{fig:02dispersion}(a), the external $\bf{B}$ breaks the $\mathcal{T}$ symmetry, and simultaneously the non-identical nanoparticles $A$ and $B$ break the $\mathcal{P}$ and $\mathcal{RT}$ symmetry, see Fig.~\ref{fig:03RT}.
The non-reciprocal band could then be obtained shown in Fig.~\ref{fig:02dispersion}(a) (ii).
In contrast, when $\mathcal{RT}$ acts on the system shown in Fig.~\ref{fig:02dispersion}(b), $\bf{B}$ is flipped twice and remains unchanged.
Meanwhile, particles $A$ and $B$ are on the $x$-axis, $\mathcal{R}$ would not modify their positions,
and thus the transformed system is identical to the non-transformed one, which means it has $\mathcal{RT}$ symmetry.
we get reciprocal dispersion relation in Fig.~\ref{fig:02dispersion}(b)(ii).

Therefore, in order to obtain nonreciprocal bands in 1D magnetized chain system, it is essential to break all related symmetries, including $\mathcal{P}$, $\mathcal{T}$ and $\mathcal{RT}$ symmetries.

%the system in Fig.~\ref{fig:02dispersion}(b) just breaks the $\mathcal{T}$ and $\mathcal{P}$ symmetry,
%while keeping the $\mathcal{RT}$ symmetry. We then see the .

\section{IV. Symmetry operators on diatomic chain system}
A system is said to have $\Theta$ symmetry if it is invariant under transformation, $\Theta^{-1}{\bf{M}}_k(\omega)\Theta={\bf{M}}_k(\omega)$. Here we show explicitly that the $\mathcal{T}$ and $\mathcal{RT}$ operator on the coupled dipole equation ${\bf{M}}_k(\omega)$. For simplicity, we consider the system shown in Fig.~\ref{fig:01}, employ quasi-static dipolar Geen's function, use simple lossless Drude model, and neglect the radiation term in polarizability.

\subsection{A. Coupled dipole equation in quasi-static limit}
Quasistatic expressions are obtained by taking free space wave vector $k_0\rightarrow0$, so the polarizability, from Eq.~(\ref{eq:a1polarizability}), is
\begin{equation}\label{eq:04quasiAlpha}
{\bm{\alpha}}'_{\sigma}{}^{ - 1} =
    \frac{1}{{{\epsilon _0}{V_\sigma }}}
    \left[ {{\bf{L}}_{\sigma} + \left( {\begin{array}{*{20}{c}}
   { - {\omega ^2}} & {i\omega {\omega _c}} & 0  \\
   { - i\omega {\omega _c}} & { - {\omega ^2}} & 0  \\
   0 & 0 & { - {\omega ^2}}  \\
\end{array}} \right)\frac{1}{{\omega _p^2}}} \right],
\end{equation}
in which we are assuming lossless model, with $\gamma=0$. Symbols are defined under Eq.~(\ref{eq:a1polarizability}). Also, the quasi-static dipolar Green's function, from Eq.~(\ref{eq:a4green}), is
\begin{equation}\label{eq:05quasiG}
{\bf{G}}'_{k\sigma\sigma'}
\equiv
\lim_{k_0\rightarrow0}
{\bf{G}}_{k\sigma\sigma'}=\frac{{1}}{{4\pi \epsilon_0}}\sum_{m\neq0}{\bf{C}}_{\sigma\sigma'}(m)\frac{{e^{ikmd}}}{{r_{0m\sigma\sigma'}^3}},
\end{equation}
where for the case shown in Fig.~\ref{fig:01}, relative position vectors ${\bf{r}}_{0m\sigma\sigma'}$ are defined by Eq.~(\ref{eq:a6withoutRT}), and
\begin{subequations}
\begin{align}
&{{\bf{C}}_{AA}}(m) = {{\bf{C}}_{BB}}(m) = \left( {\begin{array}{*{20}{c}}
   2 & 0 & 0  \\
   0 & { - 1} & 0  \\
   0 & 0 & { - 1}  \\
\end{array}} \right)\label{eq:06quasiC1}\\
&{{\bf{C}}_{AB}}(m) = \left( {\begin{array}{*{20}{c}}
   \frac{{3m^{2}d^{2}}}{{m^{2} d^{2}+t_y^2}}-1 & \frac{{3mdt_y}}{{m^{2}d^{2}+t_y^2}} & 0  \\
   \frac{{3mdt_y}}{{m^{2}d^{2}+t_y^2}} & \frac{{3t_y^{2}}}{{m^{2} d^{2}+t_y^2}}-1 & 0  \\
   0 & 0 & { - 1}  \\
\end{array}} \right)\label{eq:06quasiC3}\\
&{\bf{C}}_{BA}(m)={\bf{C}}_{AB}(-m) \label{eq:06quasiC2}
\end{align}
\end{subequations}
The above was obtained by putting Eq.~(\ref{eq:a6withoutRT}) into (\ref{eq:a5C}) in appendix B. Some properties of ${\bf{G}}'_{k\sigma\sigma'}$ are discussed in appendix D.

\subsection{B. $\mathcal{T}$ and $\mathcal{RT}$ transformation on the system}
%T operator%
When $\mathcal{T}$ operates on the system, it turns $k$ into $-k$ and takes complex conjugate (see also appendix E):
\begin{equation}\label{eq:08TMT}
T^{-1}{\bf{M}}_k(\omega)T={\bf{M}}_{-k}(\omega)^*.
\end{equation}
In the quasi-static limit, Eq.~(\ref{eq:08TMT}) can be written as
%\begin{subequations}
\begin{equation}\label{eq:08TaT}
{\mathcal{T}}^{-1}({\bm{\alpha}}'_{\sigma}{}^{-1}-{\bf{G}}'_{k\sigma\sigma'}){\mathcal{T}}=({\bm{\alpha}}'_{\sigma}{}^{-1})^*-{\bf{G}}'_{k\sigma\sigma'},
\end{equation}
%and by Eq.~(\ref{eq:08gk}),
%\begin{equation}\label{eq:08TGT}
%{\mathcal{T}}^{-1}{\bf{G}}'_{k\sigma\sigma'}{\mathcal{T}}={\bf{G}}'_{k\sigma\sigma'}.
%\end{equation}
%\end{subequations}
Therefore, ${\mathcal{T}}^{-1}{\bf{M}}_k(\omega){\mathcal{T}}={\bf{M}}_k(\omega)$ only if $({\bm{\alpha}}'_{\sigma}{}^{-1})^*={\bm{\alpha}}'_{\sigma}{}^{-1}$, which means $\omega_c=0$. That is, the system has $\mathcal{T}$ symmetry if external ${\bf{B}}=0$.

%RToperator%
For the ${\mathcal{RT}}$ transformation, first we notice that $R^{-1}=R$, then $({\mathcal{RT}})^{-1}{\bf{M}}_k(\omega){\mathcal{RT}}={\mathcal{TR}}{\bf{M}}_k(\omega){\mathcal{RT}}$. In the quasi-static limit for the case shown in Fig.~\ref{fig:01}, we have
\begin{subequations}\label{eq:10RTM}
\begin{equation}
\begin{aligned}
 &{\mathcal{TR}}{\bm{\alpha}}'_\sigma{}^{ - 1}{\delta _{\sigma \sigma '}}{\mathcal{RT}} \\
 &= {\mathcal{T}}\left[ {\begin{array}{*{20}{c}}
   0 & {\bf{R}}  \\
   {\bf{R}} & 0  \\
\end{array}} \right]\left[ {\begin{array}{*{20}{c}}
   {\bm{\alpha}}'_A{}^{ - 1} & 0  \\
   0 & {\bm{\alpha}}'_B{}^{ - 1}  \\
\end{array}} \right]\left[ {\begin{array}{*{20}{c}}
   0 & {\bf{R}}  \\
   {\bf{R}} & 0  \\
\end{array}} \right]{\mathcal{T}}\\
&= {\mathcal{T}}\left[ {\begin{array}{*{20}{c}}
   {\bm{\alpha}}'_B{}^{ - 1*} & 0  \\
   0 & {\bm{\alpha}}'_A{}^{ - 1*}  \\
\end{array}} \right]{\mathcal{T}}
  = \left[ {\begin{array}{*{20}{c}}
   {\bm{\alpha}}'_B{}^{ - 1} & 0  \\
   0 & {\bm{\alpha}}'_A{}^{ - 1}  \\
\end{array}} \right].
\end{aligned}
\end{equation}
and similarly,
\begin{equation}
{\mathcal{TR}}{\bf{G}}'_{k\sigma \sigma '}{\mathcal{RT}}={\bf{G}}'_{k\sigma \sigma '}.\\
\end{equation}
\end{subequations}
%Similarly, using Eq.~(\ref{eq:07RTfig1}), (\ref{eq:09Gproperties1}), (\ref{eq:09Gproperties2}), and (\ref{eq:08TGT}),
%\begin{equation}
%\begin{aligned}
% &{\mathcal{TR}}{\bf{G}}'_{k\sigma \sigma '}{\mathcal{RT}} \\
% &= T\left[ {\begin{array}{*{20}{c}}
%   0 & {\bf{R}}  \\
%   {\bf{R}} & 0  \\
%\end{array}} \right]\left[ {\begin{array}{*{20}{c}}
%   {{\bf{G}}'_{kAA}} & {{\bf{G}}'_{kAB}}  \\
%   {{\bf{G}}'_{kBA}} & {{\bf{G}}'_{kBB}}  \\
%\end{array}} \right]\left[ {\begin{array}{*{20}{c}}
%   0 & {\bf{R}}  \\
%   {\bf{R}} & 0  \\
%\end{array}} \right]{\mathcal{T}} \\
% &= {\mathcal{T}}\left[ {\begin{array}{*{20}{c}}
%   {{\bf{RG}}'_{kBB}{\bf{R}}} & {{\bf{RG}}'_{kBA}{\bf{R}}}  \\
%   {{\bf{RG}}'_{kAB}{\bf{R}}} & {{\bf{RG}}'_{kAA}{\bf{R}}}  \\
%\end{array}} \right]{\mathcal{T}} \\
%&=\left[ {\begin{array}{*{20}{c}}
%   {{\bf{G}}'_{kAA}} & {{\bf{G}}'_{kAB}}  \\
%   {{\bf{G}}'_{kBA}} & {{\bf{G}}'_{kBB}}  \\
%\end{array}} \right]
% = {\bf{G}}'_{k\sigma \sigma '}. \\
%\end{aligned}
%\end{equation}
%\end{subequations}
Eq.~(\ref{eq:10RTM}) implies that the $\mathcal{RT}$ transformed system is not generally identical to the non-transformed one. So the chain in Fig.~\ref{fig:01} has $\mathcal{RT}$ symmetry only if ${\bm{\alpha}}'_A={\bm{\alpha}}'_B$, which is not true.

The dispersion relation can be obtained by solving determinant equation $\det\left[{\bf{M}}_k(\omega)\right]=0$. Neglecting $z$ components,
Appendix D shows that it is a polynomial with $G'_{kAB,xy}$ up to 2nd order, and $G'_{kAB,xy}$ is the only term which is odd in $k$ in the determinant polynomial. Non-reciprocal dispersion only comes out when the polynomial is not an even function of $k$, which means coefficient of $G'_{kAB,xy}$, given by
\begin{equation*}
\begin{aligned}
&\frac{{-2i\omega {\omega_c}}}{{\omega _p^2{\epsilon _0}}}\left( {\frac{1}{{{V_A}}} - \frac{1}{{{V_B}}}} \right)\\
&\quad\times\left( {{G'_{kAA,xx}}{G'_{kAB,yy}} - {G'_{kAA,yy}}{G'_{kAB,xx}}} \right),
\end{aligned}
\end{equation*}
is non-zero. It is the case that both $\omega_c \neq 0$ and $V_A \neq V_B$, that is, external magnetic field ${\bf{B}}\neq 0$ and the particles $A$ and $B$ are not identical. Therefore, $\mathcal{T}$ and $\mathcal{RT}$ symmetry should not be present in our case of Fig. 1.

\section{V. One-way wave propagation and Energy transmission}
The non-reciprocal bands in Fig. \ref{fig:02dispersion}(a)(ii) predicts only guided modes with positive group velocities are allowed within the range $\Delta \omega$, which gives one-way propagation behavior. We demonstrate the propagation behavior and energy transmission of the magnetized diatomic chain by considering it's finite version in this section.

\subsection{A. One-way wave propagation}
\begin{figure}[htbp]
\centering
\includegraphics[width=2.6in]{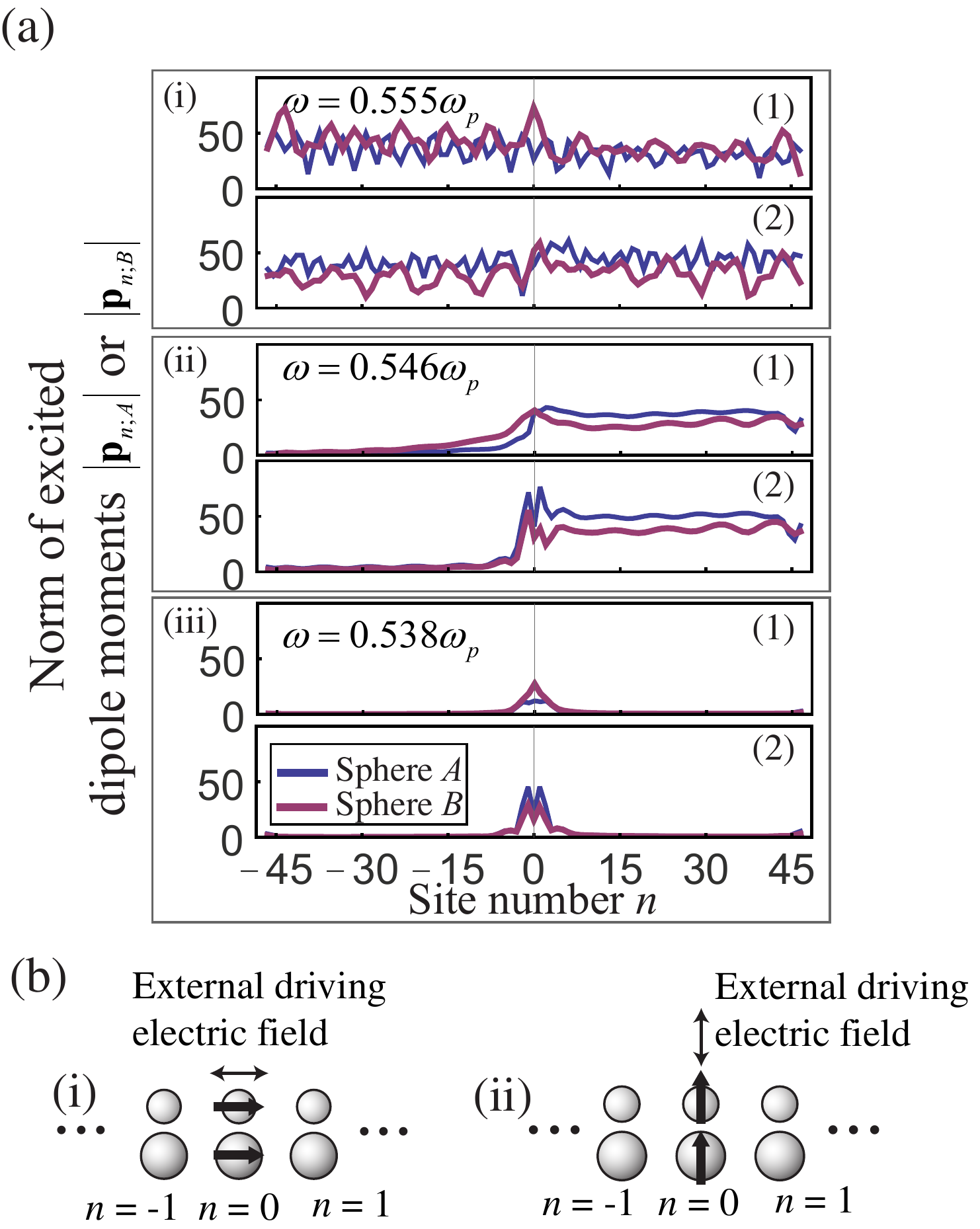}
\caption{\small(Color online) (a) Dipole moments excited on a finite magnetized diatomic chain system (93 cells). An external driving electric field is applied to the middle of the chain ($n=0$) with driving frequency (i) $\omega=0.555\omega_p$, (ii) $0.546\omega_p$ , (iii) $0.538\omega_p$, are indicated by dotted lines in Fig.~\ref{fig:02dispersion}(a). The sub-figures in each frequency are corresponding to two types of driving polarization, coherent in $x$ or $y$ direction, labelled by (1) and (2). Note that at $\omega=0.546\omega_p$, one-way propagation occurred, as only modes with positive $k$ are allowed. (b) External driving electric polarization. (i) coherent in $x$ direction; (ii) coherent in $y$ direction.}
\label{fig:04}
\end{figure}

For a finite magnetized diatomic chain system containing $N$ unit cells, Eq.~(\ref{eq:01coupled}) can be written in a matrix equation form ${\bf{M}}(\omega){\bf{p}}={\bf{E}}^{\rm{ext}}$, where ${\bf{M}}(\omega)$ is a $\left(12N+6\right)\times\left(12N+6\right)$ square matrix vectorized from ${\mathbf{M}}_{nm\sigma\sigma'}={\bm{\alpha}}^{-1}_{\sigma}\left(\omega\right)\delta_{nm}\delta_{\sigma\sigma'}
-{\mathbf{G}}_{nm\sigma\sigma'}$. ${\bf{p}}$ and ${\bf{E}}^{\rm{ext}}$ are column vectors vectorized from ${\bf{p}}_{m;\sigma'}$ and ${\bf{E}}_{n;\sigma}^{\rm{ext}}$, where each has $(2N+1)\times2\times 3=12N+6$ elements ($2N+1$ unit cells, 2 atoms per unit cell, and 3 spacial dimensions in our system). Since $\bf{M}(\omega)$ is known, and ${\bf{E}}^{\rm{ext}}$ depends on our choices, so the excited dipole moments $\bf{p}$ can be found by evaluating the inverse:
\begin{equation}\label{eq:dipoleExcited}
\bf{p}=\bf{M}(\omega)^{-1}\bf{E}^{\rm{ext}}.
\end{equation}
We study the finite version of magnetized diatomic chain in Fig.~\ref{fig:02dispersion}(a), with $N=46$ and  plasma collision frequency $\gamma=0$. Two types of driving polarizations are applied only to the site $n=0$, coherent in $x$ direction, or coherent in $y$ direction, see Fig.~\ref{fig:04}(b)(i) and (ii). In both cases, ${\bf{E}}_{n;\sigma}^{\rm{ext}}=0$ for $n\neq0$, while ${\bf{E}}_{0;A}^{\rm{ext}}={\bf{E}}_{0;B}^{\rm{ext}}=\left(1,0,0\right)^{\rm{T}}$ for case (i), and ${\bf{E}}_{0;A}^{\rm{ext}}={\bf{E}}_{0;B}^{\rm{ext}}=\left(0,1,0\right)^{\rm{T}}$ for case (ii).

The norm of excited dipole moments by three driving frequencies are shown in Fig.~\ref{fig:04}(a). Sub-figures (1) and (2) correspond to coherent driving in $x$ direction and $y$ direction. From Fig.~\ref{fig:04}(a), we see the system supports two-way propagation at $\omega=0.555\omega_p$, as there are excitations throughout the chain; and supports one-way propagation at $\omega=0.546\omega_p$, as only spheres on the right are excited; and no supported modes at $\omega=0.538\omega_p$, as there is no excitation on both sides. It is reasonable, as slope of the dispersion relation is group velocity of the coupled plasmon mode. From Fig.~\ref{fig:02dispersion}(a), we see at $\omega=0.546\omega_p$, only mode with positive $k$ is allowed, therefore only mode with $+x$ propagation is supported at this frequency. Note that reversing the propagation direction can be easily done by just flipping the direction of the external static magnetic field, this gives us a switchable optical isolator.

\subsection{B. One-way energy transmission}
\begin{figure}[htbp]
\centering
\includegraphics[width=3in]{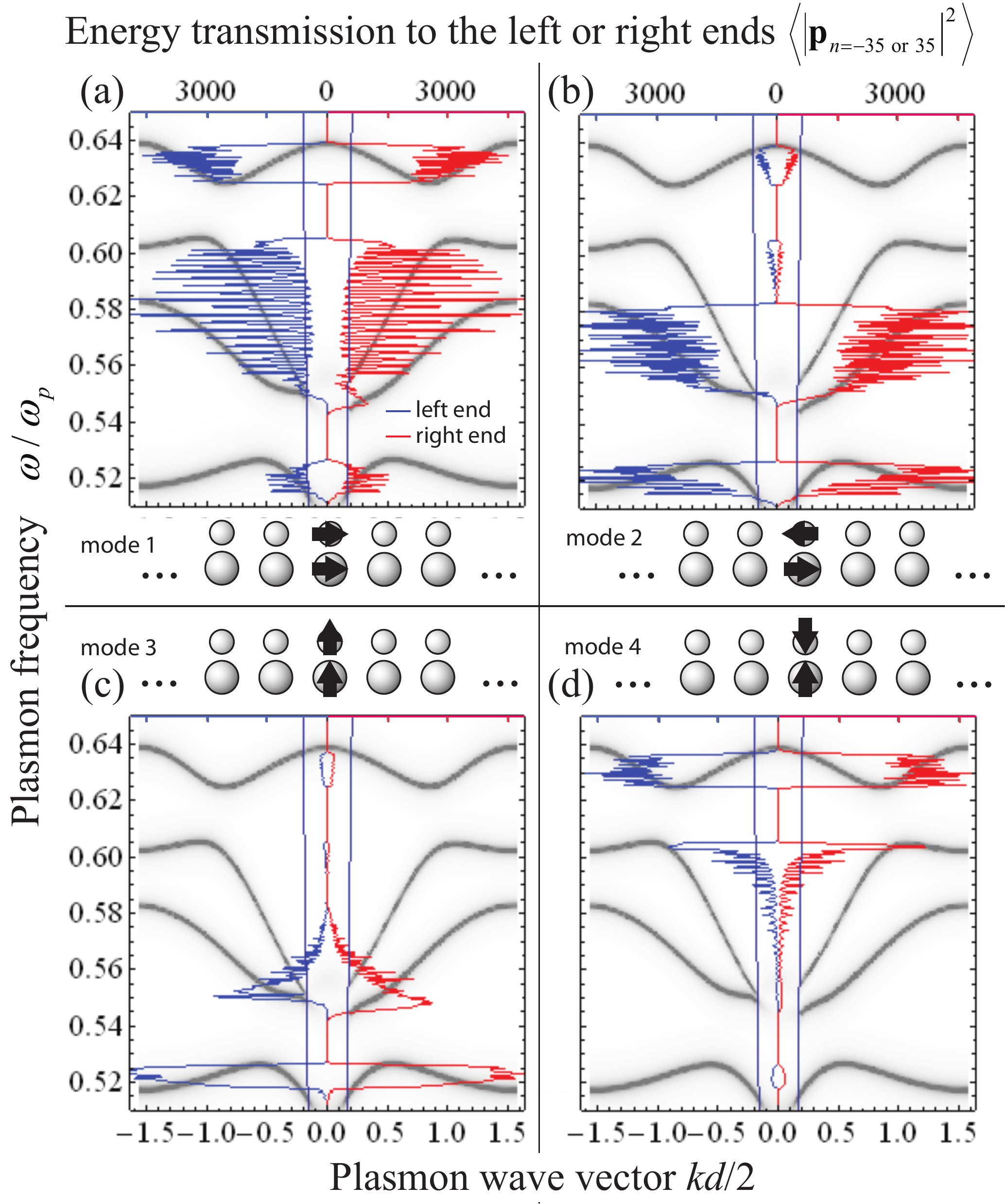}
\caption{\small(Color online) Energy transmissions to the left end ($n=-35$) or the right end ($n=35$). Energy density is proportional to the quantity $\left\langle {{{| {{{\bf{p}}_n}} |}^2}} \right\rangle$ (defined in text), and is plotted in the figure. Left (right) end quantity is denoted by the blue (red) curve and uses the left (right) upper horizontal frame ticks. The magnetized diatomic chain dispersion relation is drawn as a background for reference. Sub-figures (a)-(d) are corresponding to different external driving polarizations, which are pictured in the sub-figures (mode 1-4). One-way propagation property is demonstrated at $\omega\sim0.546\omega_p$, at where red curve is finite but blue curve is zero. A little material damping $\gamma=0.0004\omega_p$ is added to the nanoparticles here, which is to reduce extreme fluctuations so that excitations can be seen clearer.}
\label{fig:05energy}
\end{figure}

%%spacial average%%
Energy transmission is usually hard to define in plasmonic waveguides. Here we infer the transmission of the diatomic chain shown in Fig.~\ref{fig:01} by reading the dipole moments ${\bf{p}}_{n;\sigma}$ excited on nanoparticles. Since energy density is proportional to square of electric field $\left|{\rm{Re}}({\bf{E}})\right|^2$, and since dipole moment satisfies ${\bf{p}}={\bm{\alpha}}{\bf{E}}$, time averaged energy density of a particle $\sigma$ at cell $n$ is thus proportional to $|{\bf{p}}_{n;\sigma}|^2\equiv{\bf{p}}_{n;\sigma}\cdot{\bf{p}}_{n;\sigma}^*$, which gives the sense of energy transmission. We consider the quantity defined by
\begin{equation*}
\left\langle {{{| {{{\bf{p}}_n}} |}^2}} \right\rangle  = \frac{1}{9}\sum\limits_{m = n - 4}^{n + 4} {\left( {{{| {{{\bf{p}}_{m;A}}}|}^2} + {{| {{{\bf{p}}_{m;B}}} |}^2}} \right)}.
\end{equation*}
In the above we picked 4 cells near the $n$th for spatial average. One can improve by picking more cells, but in that case the chain has to be longer, or this is no longer a local quantity at cell $n$. This quantity is displayed in Fig.~\ref{fig:05energy}.
%%4 polarizations
There are 4 driving polarization modes to excite the system, two in phase and two out of phase \footnote{For mode 1, ${\bf{E}}_{0;A}^{\rm{ext}}={\bf{E}}_{0;B}^{\rm{ext}}=(1,0,0)^{\rm{T}}$; for mode 2, ${\bf{E}}_{0;A}^{\rm{ext}}=(1,0,0)^{\rm{T}}$ and ${\bf{E}}_{0;B}^{\rm{ext}}=(-1,0,0)^{\rm{T}}$; for mode 3, ${\bf{E}}_{0;A}^{\rm{ext}}=(0,1,0)^{\rm{T}}$ and ${\bf{E}}_{0;B}^{\rm{ext}}=(0,-1,0)^{\rm{T}}$; for mode 4, ${\bf{E}}_{0;A}^{\rm{ext}}=(0,1,0)^{\rm{T}}$ and ${\bf{E}}_{0;B}^{\rm{ext}}=(0,-1,0)^{\rm{T}}$.}, denoted by mode 1 to 4 as shown in the sub-figures of Fig.~\ref{fig:05energy}. It shows that there is no energy transmission within band gaps, since $\left\langle | {\bf{p}}_{35} |^2 \right\rangle = \left\langle | {\bf{p}}_{-35} |^2 \right\rangle =0$ at $\omega\sim0.54\omega_p$ and $\omega\sim0.61\omega_p$. The one-way property can be seen at $\omega=0.546\omega_p$ in Fig.~\ref{fig:05energy} (a)-(c), in which red curve is finite while blue curve is zero, implying there is energy transmission to the right end but no transmission to the left end.

From Fig.~\ref{fig:05energy} we also see that different modes excite different frequency range. Polarization modes 1 to 4 correspond to $\omega>0.55\omega_p$, $\omega<0.58\omega_p$, $\omega<0.56\omega_p$, and $\omega>0.62\omega_p$ with respectively. This can be explained by using band hybridization model (discussed in next section), which is shown in Fig.~\ref{fig:02dispersion}(a)(ii).

\section{VI. Hybridization of bands in diatomic chain}

\begin{figure}[htbp]
\centering
\includegraphics[width=3in]{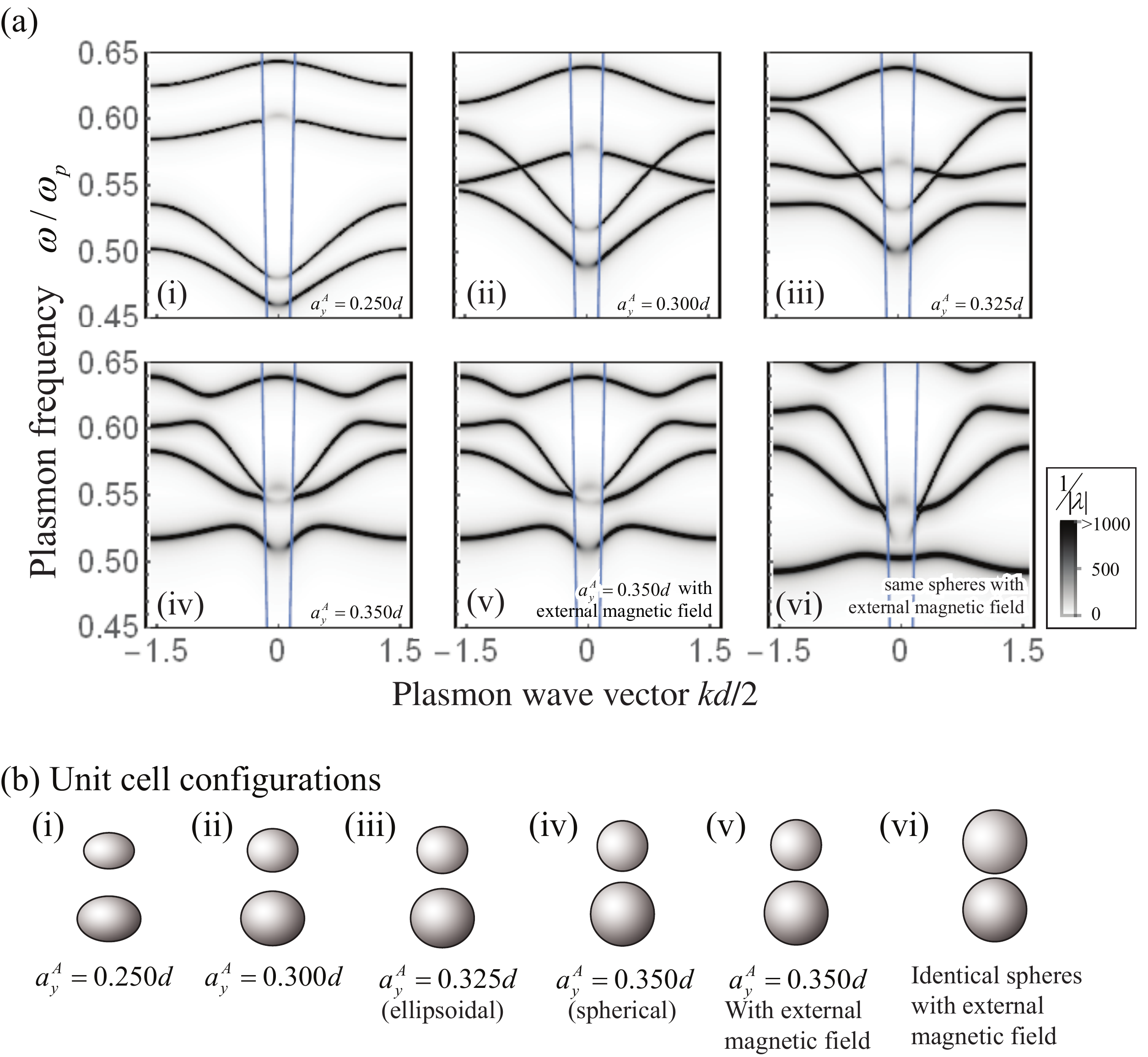}
\caption{\small(a) Dispersion relations of diatomic chains formed by different ellipsoids. Corresponding unit cell configurations are shown in (b). Moving from (i) to (vi), semi-minor axis $a_y^{\sigma}$ are increased one by one, so nanoparticles are becoming more and more spherical. ${\bf{B}}=0$, $a_x^A=0.35d$, and $V_B/V_A=0.5$ in (i) to (vi). From (i) to (vi), the lower two bands move upwards, cross the upper two bands, and hybridize each other. Note that (iv) is the case that $A$ and $B$ are spherical. (v) is different from (iv) by adding external static magnetic field ($\omega_c=0.005\omega_p$), resulting a non-reciprocal dispersion. In (vi), $A$ and $B$ are identical spheres with $a_x^A=a_x^B=0.35d$. Although ${\bf{B}}\neq0$, the dispersion is reciprocal as it is protected by $\mathcal{P}$ symmetry.}
\label{fig:06ellipsoids}
\end{figure}

There are four surface plasmon resonant modes for isolated dimer particles, two transverse modes (inphase and antiphase oscillations) and two longitudinal modes \cite{viktor2008}. The four modes form dispersion bands when the dimer particles are duplicated, becoming a diatomic chain. After then, the dispersion relation of the magnetized diatomic chain can be understood as a result of four bands hybridization, colored in blue, yellow, red, and green in Fig.~\ref{fig:02dispersion}(a)(ii). We can see the transition of the four bands in Fig.~\ref{fig:06ellipsoids}.

For better understanding of the hybridization, here we consider the diatomic chain formed by ellipsoids. Spheres in Fig.~\ref{fig:01} are replaced by ellipsoids with varying $a_x^{\sigma}$ and $a_y^{\sigma}$, where $t_y=0.75d$. Fig.~\ref{fig:06ellipsoids}(a)(i)-(iv) show the dispersion relations with increasing $a_y^{\sigma}$. In these 4 cases, $a_x^A=0.35d$, $V_B/V_A=0.5$, and ${\bf{B}}=0$. Corresponding unit cell structures are depicted in (b)(i) to (b)(iv). Fig.~\ref{fig:06ellipsoids} (a)(i) shows four separated bands. The lower two bands move up and cross the upper two bands when the particles are more spherical, as shown in (i)-(iv). If one further applies external static magnetic field to case (iv) such that $\omega_c=0.005\omega_p$, we have case (v), which is the case in Fig.~\ref{fig:02dispersion}(a). Thus, in case (i)-(v), we see the deformation of bands, and so conclude that the dispersion in Fig.~\ref{fig:06ellipsoids}(iv) is formed by hybridization of the four bands. Case in Fig.~\ref{fig:06ellipsoids}(b)(vi) shows the case with nanoparticle $B$ is identical to nanoparticle $A$, that is $a_x^{\sigma}=a_y^{\sigma}=0.35d$, where $\sigma=A$ or $B$. This is obtained by replacing all particle $B$ by $A$ in Fig.~\ref{fig:06ellipsoids}(b)(v). Bands in Fig.~\ref{fig:06ellipsoids}(b)(vi) are symmetric, as it has both $\mathcal{RT}$ and $\mathcal{P}$ symmetry. From these we know the exitance of $\bf{B}$ and non-identical spheres are essential to achieve non-reciprocal bands.

The hybridization model also explains the excitations in Fig.~\ref{fig:05energy}. Fig.~\ref{fig:02dispersion}(a)(ii) shows the upper two bands contain parts that are original from the red dashed band, therefore the excitation by polarization mode 1 is prominent at the upper two bands, as found in Fig.~\ref{fig:05energy}(a). This is because the oscillation mode of the red dashed band at $k=0$ is the same as mode 1. Similarly, the lower two bands contain parts that are original from the green dashed band, and hence excitation by polarization mode 2 is prominent at the lower two bands.

\section{Conclusions}
%advantage and one way
%%%
%There are many schemes to achieve optical isolators. When comparing to some of the others, our model has a smaller requirement on external magnetic field ${\bf{B}}$\footnote{$\bf{B}$ is limited by experimental realization, and thus cannot be too large. Further discussions on the order of $\bf{B}$ can be found in \cite{zongfuyu2013}.}, where cyclotron frequency $\omega_c\sim0.005\omega_p$ is an order smaller than that used in \cite{yhadad2010,achristofi2013}, and two order smaller than that in \cite{zongfuyu2013}. We also have a wider operation frequency range than that in \cite{yhadad2010}, where frequency gap is about $1\times10^{-4}\omega_p$ in \cite{yhadad2010}, and is about $1\times10^{-2}\omega_p$ in our case.
%%%

To conclude, we used a compact non-chiral magnetized plasmonic chain to demonstrate the crucial role of $\mathcal{RT}$ symmetry in the design of these subwavelength waveguides with non-reciprocal dispersion ($\omega(-k)\neq\omega(k)$). The hybridization among four plasmon modes and free photon modes give rise to a frequency range where only guided modes in one direction are allowed. While we are considering a weaker magnetic field, this operation frequency range is already much wider than that in Ref. \cite{yhadad2010}. Matrix representations of the symmetry operators were used to explain explicitly how the spectral reciprocity is protected by $\mathcal{T}$, $\mathcal{P}$, and $\mathcal{RT}$ symmetries. The results may provide a clear guidance on the design of one-way waveguides.

\section{acknowledgement}
This work was supported by the Hong Kong Research Grant Council through the Area of Excellence Scheme (grant no. AoE/P-02/12) and the Hong Kong Polytechnic University under grant no. G-YBCH. We thank Prof. C. T. Chan and Dr. S. W. Su for useful discussions.

\renewcommand{\theequation}{A-\arabic{equation}}
% redefine the command that creates the equation no.
\setcounter{equation}{0}  % reset counter
\section*{APPENDIX}  % use *-form to suppress numbering

\subsection{A. Polarizability and Drude model in external $\bf{B}$}
The inverse polarizability of an ellipsoidal particle $\sigma$ ($a_x^{\sigma}\geq a_y^{\sigma}=a_z^{\sigma}$) is given by \cite{yhadad2010,AS1999,BCF1983}
\begin{equation}\label{eq:a1polarizability}
{{\bm{\alpha}}_{\sigma}^{ - 1}} = \frac{1}{{{\epsilon _0}V}}\left[ {{{\left( {{\frac{{{\bm{\epsilon }(\omega)}}}{{\epsilon_0}}} - {\bf{I}}} \right)}^{ - 1}} + {\bf{L}}_{\sigma}} \right]-\frac{{ik_0^3}}{{6\pi\epsilon_0}}{\bf{I}},
\end{equation}
where volume $V_{\sigma}\equiv 4\pi a_x^{\sigma} a_y^{\sigma} a_z^{\sigma}/3$, ${\bf{I}}$ is identity $3\times3$ matrix, $k_0$ is light wave vector in free space, and parameter ${\bf{L}}_{\sigma}$ depends on particle's shape, which is given by ${\bf{L}}_{\sigma}={\text{diag}}(N_x,N_y,N_z)$. For spherical particles, $N_x=N_y=N_z=1/3$. For ellipsoidal particles, $N_x=(1+e^2)[{\ln{\frac{{1+e}}{{1-e}}}-2e}]/(2e^3)$ and $N_y=N_z=(1-N_x)/2$, where $e={\sqrt{1-{a_y^{\sigma}}^2/{a_x^{\sigma}}^2}}$. As we consider nanoparticles $A$ and $B$ are of the same shape in this paper, so ${\bf{L}}_A={\bf{L}}_B$. The last term with $k_0^3$ accounts for the radiation correction \cite{Albaladejo2010}, which will be vanished in quasi-static approximation.

With Drude model, the dielectric tensor is in the form \cite{hfzhang2012,theoryMetamaterials}
\begin{equation}\label{eq:a2Drude}
    \frac{{{\bm{\epsilon}} (\omega )}}{{{\epsilon _0}}} =
        \left( {\begin{array}{*{20}{c}}
        {{\epsilon _{xx}}} & {{\epsilon _{xy}}} & 0  \\
        { - {\epsilon _{xy}}} & {{\epsilon _{yy}}} & 0  \\
        0 & 0 & {{\epsilon _{zz}}}  \\
    \end{array}} \right),
\end{equation}
in which $\epsilon_{xx}=\epsilon_{yy}=1-\omega_p^2(\omega+i\gamma)/[\omega(\omega+i\gamma)^2-\omega\omega_c^2)]$, $\epsilon_{xy}=-i\omega_p^2\omega_c/[\omega(\omega+i\gamma)^2-\omega\omega_c^2]$, $\epsilon_{zz}=1-\omega_p^2/(\omega^2+i\gamma\omega)$. $\gamma$, $\omega_p$, and $\omega_c$ are plasma collision frequency, plasmon frequency, and cyclotron frequency. In the case without material loss, $\lambda=0$, and ${\bm{\epsilon}}(\omega)$ becomes real. More terms have to be added to the static polarizability if one wants an accurate numerical evaluation on lossy materials, but it is not of interest in this paper. Further discussions can be found in Ch.8 in \cite{theoryMetamaterials}.

\subsection{B. Coupled dipole equation in $k$ space}
In Eq.~(\ref{eq:01coupled}), defining ${\bf{M}}_{nm\sigma\sigma'}={\bm{\alpha}}^{-1}_{\sigma}\delta_{nm}\delta_{\sigma\sigma'}-{\bf{G}}_{nm\sigma\sigma'}$ , putting ${\bf{p}}_{m;\sigma'}={\bf{p}}_{k;\sigma'}e^{ikmd}$, and in the case ${\bf{E}}_{n;\sigma}^{\rm{ext}}=0$, we have
\begin{align}
\sum\nolimits_{\sigma'}\left(\sum\nolimits_{m}{{\bf{M}}_{nm\sigma\sigma'}e^{ikmd}}\right){\bf{p}}_{k;\sigma'}=0. \nonumber
\end{align}
Multiplying both sides by $e^{-iknd}$, and notice ${\bf{M}}_{nm\sigma\sigma'}$ depends on $m-n$ only, we have ${\bf{M}}_{nm\sigma\sigma'}={\bf{M}}_{0,m-n,\sigma\sigma'}$, and thus
\begin{align}
\sum\nolimits_{\sigma'}\left(\sum\nolimits_{m}{{\bf{M}}_{0,m-n,\sigma\sigma'}e^{ik(m-n)d}}\right){\bf{p}}_{k;\sigma'}&=0. \nonumber\end{align}
As the sum $m$ runs from $-\infty$ to $\infty$, we have
\begin{align}
\sum\nolimits_{\sigma'}\left(\sum\nolimits_{m}{{\bf{M}}_{0m\sigma\sigma'}e^{ikmd}}\right){\bf{p}}_{k;\sigma'}&=0. \nonumber
\end{align}
Writing it in matrix form, we have
\begin{multline}\label{eq:a3coupled}
\sum\limits_{m} {\left\{ {\left[ {\begin{array}{*{20}{c}}
   {{\bm{\alpha}} _A^{ - 1}} & 0  \\
   0 & {{\bm{\alpha}} _B^{ - 1}}  \\
\end{array}} \right]\delta _{0m}  } \right.}
\\
\left.
{\begin{array}{*{20}{c}}
\\
\\
\end{array}}
- \left[ {\begin{array}{*{20}{c}}
   {{{\bf{G}}_{0mAA}}} & {{{\bf{G}}_{0mAB}}}  \\
   {{{\bf{G}}_{0mBA}}} & {{{\bf{G}}_{0mBB}}}  \\
\end{array}}\right]
{e^{ikmd}}\right\}
\left[ {\begin{array}{*{20}{c}}
   {{{\bf{p}}_{k;A}}}  \\
   {{{\bf{p}}_{k;B}}}  \\
\end{array}} \right] = 0.
\end{multline}
This gives Eq.~(\ref{eq:02matrixcouple}).

\subsection{C. Dynamic dipolar Green function for diatomic chain system}
The dipolar coupling between the particle $\sigma'$ in the $m$th cell and the particle $\sigma$ in the $n$th cell depends on $m-n$ only, i.e., ${\bf{G}}_{nm\sigma\sigma'}={\bf{G}}_{0,m-n,\sigma\sigma'}$, so we only show elements ${\bf{G}}_{0m\sigma\sigma'}$ here. Also, position vector of the particle $\sigma'$ in the $m$th cell is denoted by ${\bf{r}}_{m\sigma'}$. Relative position vector is then ${\bf{r}}_{0m\sigma\sigma'}\equiv{\bf{r}}_{0\sigma}-{\bf{r}}_{m\sigma'}$, and the corresponding unit vector is therefore ${\bm{\rho}}_{\sigma\sigma'}(m)\equiv{\bf{r}}_{0m\sigma\sigma'}/r_{0m\sigma\sigma'}$. Spatial components of the unit vector are denoted by $\rho_{\sigma\sigma'}(m)_x$, $\rho_{\sigma\sigma'}(m)_y$, and $\rho_{\sigma\sigma'}(m)_z$. The dynamic coupling is well known and is given by \cite{yhadad2010}
\begin{equation}\label{eq:a4green}
\begin{aligned}
    {\bf{G}}_{0m\sigma\sigma'}
    = \frac{{{e^{i{k_0}{r_{0m\sigma\sigma' }}}}}}{{4\pi {\epsilon _0}}} \times
    \left[ {{\bf{A}}_{\sigma\sigma'}(m)\frac{{k_0^2}}{{{r_{0m\sigma\sigma'}}}}} \right. \\
    \left. { + {\bf{C}}_{\sigma\sigma'}(m)\left( {\frac{{1}}{{{r_{0m\sigma\sigma'}^3}} }- \frac{{i{k_0}}}{{r_{0m\sigma\sigma' }^2}} }\right)} \right]
\end{aligned}
\end{equation}
for $m\neq 0$ together with $\sigma'\neq \sigma$, otherwise ${\bf{G}}_{00\sigma\sigma}=0$, as a particle is not interacting itself by generating electric field. In the above, $\epsilon_0$ is free space permittivity, and $k_0=\omega/c$ is the light wave vector in free space. Matrixes in the above are
\begin{equation*}
\begin{aligned}
&{\bf{A}}_{\sigma\sigma'}(m) = \\
&\left( {\begin{array}{*{20}{c}}
   {{\rho_{\sigma\sigma' }}{(m)}_y^2} & { - {\rho_{\sigma\sigma' }}{{(m)}_y}{\rho_{\sigma\sigma' }}{{(m)}_x}} & 0  \\
   { - {\rho_{\sigma\sigma' }}{{(m)}_x}{\rho_{\sigma\sigma'}}{{(m)}_y}} & {{\rho_{\sigma\sigma' }}{(m)}_x^2} & 0  \\
   0 & 0 & 1  \\
\end{array}} \right)
\end{aligned}
\end{equation*}
and
\begin{equation}\label{eq:a5C}
\begin{aligned}
&{\bf{C}}_{\sigma\sigma'}(m) = \\
&\left( {\begin{array}{*{20}{c}}
   {3{\rho_{\sigma\sigma' }}{(m)}_x^2 - 1} & {3{\rho_{\sigma\sigma' }}{{(m)}_x}{\rho_{\sigma\sigma ' }}{{(m)}_y}} & 0  \\
   {3{\rho_{\sigma\sigma' }}{{(m)}_y}{\rho_{\sigma\sigma' }}{{(m)}_x}} & {3{\rho_{\sigma\sigma' }}{(m)}_y^2 - 1} & 0  \\
   0 & 0 & { - 1}  \\
\end{array}} \right).
\end{aligned}
\end{equation}
For the system shown in Fig.~\ref{fig:01} or \ref{fig:02dispersion}(a)(i), relative position vectors are
\begin{equation}\label{eq:a6withoutRT}
\begin{split}
{{\bf{r}}_{0mAA}} = {{\bf{r}}_{0mBB}} = \left(-md,0,0\right)^{\rm{T}},\\
 {{\bf{r}}_{0mAB}} = \left(-md,-t_y,0\right)^{\rm{T}},\\
{\rm{and}}\quad
 {{\bf{r}}_{0mBA}} = \left(-md,t_y,0\right)^{\rm{T}}.
\end{split}
\end{equation}
For the system shown in Fig.~\ref{fig:02dispersion}(b)(i), relative position vectors are
\begin{equation}\label{eq:a6witRT}
\begin{split}
{{\bf{r}}_{0mAA}} = {{\bf{r}}_{0mBB}} = \left(-md,0,0\right)^{\rm{T}},\\
 {{\bf{r}}_{0mAB}} = \left(-md-t_x,0,0\right)^{\rm{T}},\\
{\rm{and}}\quad
 {{\bf{r}}_{0mBA}} = \left(-md+t_x,0,0\right)^{\rm{T}}.
\end{split}
\end{equation}

\subsection{D. Properties of quasi-static Green function ${\bf{G}}'_{k\sigma\sigma'}$ for system in Fig.~\ref{fig:02dispersion}(a)}
There are some properties about the matrix ${\bf{G}}'_{k\sigma\sigma'}$ regarding to the system in Fig.~\ref{fig:01}. From Eq.~(\ref{eq:05quasiG}), one can see
\begin{equation}\label{eq:08gk}
{\bf{G}}'_{-k\sigma\sigma'}={\bf{G}}'_{k\sigma\sigma'}{}^*.
\end{equation}
Also, from Eq.~(\ref{eq:06quasiC1}) we know ${\bf{G}}'_{kAA}={\bf{G}}'_{kBB}$, and they are both diagonal, so we have
\begin{subequations}\label{eq:09Gproperties}
\begin{equation}\label{eq:09Gproperties1}
\begin{aligned}
{\bf{R}}^{-1}{\bf{G}}'_{kAA}{\bf{R}}={\bf{G}}'_{kAA}={\bf{G}}'_{kBB}\\
{\bf{R}}^{-1}{\bf{G}}'_{kBB}{\bf{R}}={\bf{G}}'_{kBB}={\bf{G}}'_{kAA},
\end{aligned}
\end{equation}
where ${\bf{R}}=\mbox{diag}(1,-1,-1)$ is the rotation matrix about $x$-axis with $180^\circ$. Furthermore, Eq.~(\ref{eq:06quasiC2}) tells us that off diagonal elements in ${\bf{G}}'_{kAB}$ and ${\bf{G}}'_{kBA}$ are purely imaginary, and Eq.~(\ref{eq:06quasiC3}) implies ${\bf{G}}_{kAB}^*={\bf{G}}_{kBA}$, so we have
\begin{equation}\label{eq:09Gproperties2}
\begin{aligned}
&{\bf{R}}^{-1}{\bf{G}}'_{kAB}{\bf{R}}={\bf{G}}'_{kAB}{}^*={\bf{G}}'_{kBA}\\
&{\bf{R}}^{-1}{\bf{G}}'_{kBA}{\bf{R}}={\bf{G}}'_{kBA}{}^*={\bf{G}}'_{kAB}.
\end{aligned}
\end{equation}
\end{subequations}

Elements in ${\bf{G}}'_{k\sigma\sigma'}$ are functions of $k$. By expanding the terms and Eq.~(\ref{eq:09Gproperties}), we see that ${{\bf{G}}'_{kAA}}$, $G'_{kAB,xx}$ and $G'_{kAB,yy}$ are purely real and even in $k$, while $G'_{kAB,xy}$ is purely imaginary and odd in $k$. Hence, we deduced that, neglecting $z$ components,
\begin{equation*}
\begin{aligned}
    &{{\bf{G}}'_{kAA}} = {{\bf{G}}'_{kBB}} =
    \left( {\begin{array}{*{20}{c}}
       {{G'_{kAA,xx}}} & 0  \\
       0 & {{G'_{kAA,yy}}}  \\
    \end{array}} \right)\\
    &{{\bf{G}}'_{kAB}} = \left( {\begin{array}{*{20}{c}}
       {{G'_{kAB,xx}}} & {{G'_{kAB,xy}}}  \\
       {{G'_{kAB,xy}}} & {{G'_{kAB,yy}}}  \\
    \end{array}} \right)\\
    &{{\bf{G}}'_{kBA}} = \left( {\begin{array}{*{20}{c}}
       {{G'_{kAB,xx}}} & -{{G'_{kAB,xy}}}  \\
       -{{G'_{kAB,xy}}} & {{G'_{kAB,yy}}}  \\
    \end{array}} \right).
\end{aligned}
\end{equation*}
Therefore, the determinant
\begin{equation*}
|{{{\bf{M}}_k}(\omega )}| = \left| {\begin{array}{*{20}{c}}
   {\bm{\alpha}} '_A{}^{ - 1}(\omega ) - {{\bf{G}}'_{kAA}} &  - {\bf{G}}'_{kAB}  \\
   { - {{\bf{G}}'_{kBA}}} & {\bm{\alpha}}'_B{}^{ - 1}(\omega ) - {\bf{G}}'_{kAA}  \\
\end{array}} \right|
\end{equation*}
is a polynomial with $G'_{kAB,xy}$ up to 2nd order.

\subsection{E. The $\mathcal{T}$ operator}
Regarding to Eq.~(\ref{eq:05Toperator}), noticing ${\mathcal{T}}^{-1}={\mathcal{T}}$, we have
\begin{equation*}
\begin{aligned}
 & \sum\limits_m^{} {{\mathcal{T}}^{-1}{{\bf{M}}_{nm}}{\mathcal{T}}{{\bf{p}}_m}}  = \sum\limits_m^{} {{\mathcal{T}}{{\bf{M}}_{nm}}{\mathcal{T}}{{\bf{p}}_k}{e^{ikmd}}} \\
 & \quad = {\mathcal{T}}\sum\limits_m^{} {{{\bf{M}}_{nm}}{\bf{p}}_k^*{e^{ - ikmd}}}  \\
  & \quad= {\mathcal{T}}{e^{ - iknd}}\sum\limits_m^{} {{e^{iknd}}{{\bf{M}}_{nm}}{\bf{p}}_k^*{e^{ - ikmd}}}  \\
  &\quad= {\mathcal{T}}\sum\limits_m^{} {{{\bf{M}}_{nm}}{e^{ - ik(m - n)d}}} {\bf{p}}_k^*{e^{ - iknd}} \\
  &\quad= {\mathcal{T}}{{\bf{M}}_{-k}}{\bf{p}}_k^*{e^{ - iknd}} = {\bf{M}}_{-k}^*{\bf{p}}_k^{}{e^{iknd}}.
\end{aligned}
\end{equation*}In the above, we used the fact that ${\bf{M}}_{nm}={\bf{M}}_{0,m-n}$ and defined ${\bf{M}}_k=\sum\nolimits_m{\bf{M}}_{0m}e^{ikmd}$. The last line proves Eq.~(\ref{eq:08TMT}).

\subsection{F. The $\mathcal{RT}$ operator on diatomic chain system in Fig.~\ref{fig:02dispersion}(b)}
Double chain in Fig.~\ref{fig:02dispersion}(b) has $\mathcal{RT}$ symmetry. Here we show that ${\bf{M}}_k(\omega)$ is commute with $\mathcal{RT}$ in quasi-static limit. By Eqs.~(\ref{eq:07RTfig4}) and (\ref{eq:08TaT}),
\begin{equation*}
\begin{aligned}
&{\left( {\mathcal{RT}} \right)^{ - 1}}{\bm{\alpha}}'_\sigma{ }^{ - 1}{\delta _{\sigma \sigma '}}\left( {RT} \right) \\
&  = {\mathcal{T}}\left[ {\begin{array}{*{20}{c}}
   {\bf{R}} & 0  \\
   0 & {\bf{R}}  \\
\end{array}} \right]\left[ {\begin{array}{*{20}{c}}
   {\bm{\alpha}}'_A{ }^{ - 1} & 0  \\
   0 & {\bm{\alpha}}'_B{ }^{ - 1}  \\
\end{array}} \right]\left[ {\begin{array}{*{20}{c}}
   {\bf{R}} & 0  \\
   0 & {\bf{R}}  \\
\end{array}} \right]{\mathcal{T}} \\
&  = \left[ {\begin{array}{*{20}{c}}
   {\bm{\alpha}}'_A{ }^{ - 1} & 0  \\
   0 & {\bm{\alpha}}'_B{ }^{ - 1}  \\
\end{array}} \right] = {\bm{\alpha}}'_\sigma{ }^{ - 1}{\delta _{\sigma \sigma '}}. \\
\end{aligned}
\end{equation*}
Also, by putting Eq.~(\ref{eq:a6witRT}) into (\ref{eq:a5C}), we see  ${\bf{C}}_{\sigma\sigma'}(m)={\rm{diag}}(2,-1,-1)$, and thus ${\bf{G}}'_{k\sigma\sigma'}$ are diagonal. As a result, ${\bf{R}}^{-1}{\bf{G}}_{k\sigma\sigma'}{\bf{R}}={\bf{G}}_{k\sigma\sigma'}$, and we have
\begin{equation*}
{\left( {{\mathcal{RT}}} \right)^{ - 1}}{{\bf{G}}_{k\sigma \sigma '}}\left( {\mathcal{RT}} \right) = {{\bf{G}}_{k\sigma \sigma '}}.
\end{equation*}
%in which we have used Eq.~(\ref{eq:08TGT}).

\subsection{G. Comments on Pi-rotation ($\mathcal{R}$) symmetry and reflection-in-$y$ (${\mathcal{P}}_y$) symmetry}
As we assumed magnetic field is in $z$-direction, we are actually considering the reciprocity of bands related to $x$ and $y$ components (abbreviated as $xy$ bands), as $z$ related components are separated. The conclusion that ``reciprocity is protected by $\mathcal{RT}$ symmetry'' thus can be narrowed to ``reciprocity is protected by $\Theta\mathcal{T}$ symmetry'', where $\Theta$ is the $x$-$y$ plane projection of $\mathcal{R}$. However, reflection-in-$y$ (${\mathcal{P}}_y$), which changes $(x,y,z)$ into $(x,-y,z)$, has the same form under the projection. Explicitly, matrix representations of the operators are,
\begin{equation*}
{\bf{R}}=
\left(
\begin{array}{ccc}
1 & 0 &0\\
0 & -1& 0\\
0&0&-1
\end{array}
\right)
\quad\mbox{and}\quad
{\bf{P}}_y=
\left(
\begin{array}{ccc}
1 & 0 &0\\
0 & -1& 0\\
0&0&1
\end{array}
\right),
\end{equation*}
from which we see they are the same if we chop away $z$ related components. As a result, $\Theta$ is corresponding to two spatial transformations, and the transformations have the same effects on $xy$ bands. As the two transformations are equivalent under the projection, so we may also arrive the conclusion that reciprocity is also protected by ${\mathcal{P}}_y{\mathcal{T}}$ symmetry \footnote{Breaking $\Theta$ symmetry is simultaneously breaking both ${\mathcal{P}}_y$ symmetry and $\mathcal{R}$ symmetry, but breaking ${\mathcal{P}}_y$ symmetry or $\mathcal{R}$ symmetry does not mean breaking the other two symmetries.}.

%\begin{figure*}[htbp]
%\centering
%\includegraphics[width=5in]{dispersion.pdf}
%\caption{\small(a)dispersion for comparison}
%\label{fig:07dispersion}
%\end{figure*}

%\begin{figure*}[htbp]
%\centering
%\includegraphics[width=2.5in]{onewaypropagation.pdf}
%\caption{\small oneway propagation}
%\label{fig:08onewaypropagation}
%\end{figure*}

\bibliography{oneway}

\end{document}